# A solution for external costs beyond negotiation and taxation

Alexandre Magno de Melo Faria[a]   Hélde A. D. Hdom[b]

[a]Federal University of Mato Grosso, Faculty of Economics. Brazil.
[b]University of Beira Interior, Portugal

## Abstract

This article aims to launch light on the limitations of the Coase and Pigou approach in the solution of externalities. After contextualizing the need for integration of ecological and economic approaches, we are introducing a new conceptual proposal complementary to conventional economic approaches. Whose process is guaranteed by a set of diffuse agents in the economy that partially reverses entropy formation and marginal external costs generated by also diffuse agents? The approach differs in six fundamentals from traditional theory and proposes a new way of examining the actions of agents capable of reducing entropy and containing part of external costs in the market economy.

## 1. Introduction

The beginning of the 21st century has been marked by transformations that are directly or indirectly interconnected through the various problems that society, and the economy and technological development, and the environment are as their main drivers of these transformations. In particular, the energy consumption by humanity to run the economy of countries (Hdom and Fuinhas, 2020). These transformations have been increasingly accentuated since the first meeting of nations at Rio 92 to discuss the relevance of sustainability policies for the world to continue its development without compromising the survival of human beings on the Planet. Since then, technological transformations through innovation policies and energy and economic efficiency have been the subject of increasingly appropriate investigation. Therefore, this entire relationship between economy, human beings, ecosystems, and the environment is an increasingly important source of analysis. In this work, we focus on interactions that the economic system is primarily responsible for, and its direct or indirect impact entails a series of general or macroeconomic problems. Thus, a new theoretical assessment approach is launched to fill a gap in this relationship mediated by positive and negative externalities.

For example, the research provides a new theoretical framework to support researchers in rotating fields such as Environmental Sciences, Environmental Economics, Ecological Economics, Energy Economics, Natural Resource Economics, and Social Sciences, not limited to this scientific area group, more to all those who make use of science to improve the understanding of the impacts of human and economic actions and activities on the life and well-being of society. That is the search for balance between environmental, social, and economic forces, whose interactions the economic system is responsible for. The method involves the precise analysis of the development of externalities in the economy and the determinants of behavior change during the process of extracting resources for manufacturing, trading, consuming, and disposing of economic goods. In this scenario, the research facilitates the understanding of the impacts, and the results are extremely relevant to the Energy, environmental, and economic nexus. Especially related to the addition of the theme of renewable energies, energy security, and the strong reduction of emissions for the Low Carbon economy. In addition, it helps to guide future empirical research in the search for answers to problems related to externalities. In other words, this work takes a different path from the conventional or unsustainable point of view of economics to improve the understanding of the reverse impacts of externalities.

Unprecedented, the world is simultaneously experiencing a global health crisis along with economic and environmental problems. There is neither consensus, and also a quick fix to stabilize global warming at 1.5 °C as detailed in the Paris Agreement. However, the COVID19 pandemic especially revealed that the loss of life far outweighs the economic benefits in the humanitarian context. Whilst, in the economic context the data showed a sharp retraction in productive activities due to the impacts of the global health crisis. At the same time, $CO_2$ emissions and other pollution factors also drastically decreased in the environmental context during the isolation peak. The reality of the Global Pandemic has forced politicians, country governments, and policymakers to rethink the sense of priority (the lives of people, citizens, or profit and economic growth). However, the open debate seems to lead to a well-known path, that is "degrowth" and the "energy transition" as a reply to tackles the problem by parts. Which of these parts is the best choice will depend largely on the benefits associated with each of the alternatives. Another problem is that if the degrowth alternative can be a way to delay the climate crisis responding to the generational effect, this alternative can mean, to some degree, not only the worsening of people's living conditions. But also, the corresponding gap between poverty and wealth, impacting the aggregate effect. Alternatively, the energy transition poses enormous challenges, but it is more real than considering a utopian hypothesis. This is:

- How much of the world's GDP needs to decrease to have a minimum level of climate and energy security?

- Furthermore, how much of this decrease can be distributed among rich, emerging, and poor countries until the complete transition from fossil energy to renewable energy?

These are pressing questions that the search for answers can help the set of results explained by modern economic models.

Our analysis proposal is based on well-known literature, i.e., we launch the proposal to investigate these phenomena through an intrinsic relationship of positive and negative externalities, drawing a parallel simultaneously. For example, pollution imposes a cost on society. That is the loss of biodiversity, climate change, and other effects that bring costs to societies. A major problem has been the depletion of natural resources, which played a double role in the development of countries (Hdom et al., 2019). This is because the depletion of natural resources causes a negative change in the natural environment and because the natural resources extracted to carry out economic production generate waste (pollution). This dual role is theorized and explained by theories such as Natural Resources and traditionally that of externality, among others linked to the investigation of the effects of the economy on the environment.

In this sense, we have divided the work into six parts, in addition to the introduction, the second part sheds light on the limitations of Coase and Pigou's approach in solving externalities. In the third part, we contextualize the need for integration of ecological and economic approaches, and finally, in the fourth part, we present a new complementary conceptual proposal. In the sixth part, the conclusion and limitations of the research are provided.

2. Background

The debate about externalities is found in "Principles of Economics" by Marshall (1890) when the author mentions a process for the formation of external economies. This approach is well established within the welfare of the Neoclassic economy, considering the ethos of economic agents which maximize their benefit in a market. Where the agents seek to decide the better strategies for their well-being through the utility that can be achieved. The individual decisions would not include positive or negative reactions to third parties, but only selfishness capable of maximizing its benefits. For economic theory "the economic model of consumer behavior is very simple: people choose the best products they can afford to pay for." On the other hand, for Alfred Marshall, a decline in the average production costs of a company may be possible obtain, when generated by economies external to the firm, but internal to the industry. (e.g., technical innovations and information exchanges) can generate positive external economies, and the greater the general development of the industry, a positive impact on third parties is generated, but it is not perceived ex-ante or ex-post to a decision.

To Marshall external economies are not generated internally in firms, but internally in the industry. The larger is the concentration and sharing of information, the bigger are the external gains. This phenomenon is often referred to as Marshall-Arrow-Romer externalities (Sahdev, 2016). There is also another perception of this phenomenon, known as, urbanization economies, which regards the decrease of operating costs derived from the spatial concentration of multiple and diverse interdependent activities. In this process, urbanization economies are external to a firm and the industry, instead generated in other related activities, yet that reduce the overall costs because of systemic interaction. Such economies are referred to as Jacob's externalities (Jacobs 1969).

The second approach that affected deeply the externalities debate arose in Cambridge, with Pigou (1920) In this work "The Economics of Welfare", Pigou develops important contributions to economic decisions to advance the welfare of a community. Among others, one of his main concerns was based on actions to raise the income of a nation. An alternative would be de the displacement of inefficient resources. Generally, industries are interested, not in the social, but instead in the private liquid product of their stocks. Regarding moving costs, self-interest will tend to generate equality in the marginal private liquid product of resources invested in different ways. However, it will not tend to generate equality in marginal liquid social products, except when marginal private liquid product and marginal social liquid product are identical. When there is a divergence between these two types of marginal liquid products, self-interest will not tend to maximize the national product; and therefore, it is expected that certain specific acts of interference in the normal economic processes do not diminish but increase the product (Pigou, 1920).

In this sense, in seeking explanations for external economies, Pigou defines the marginal social liquid product and marginal private liquid product. Maximizing national dividends could only be achieved when there was an identity between the two marginal liquids products. In the presence of differences, Pigou believed in the existence of externalities that would affect the economic agents' ability to achieve maximization, even in competitive markets. The solutions for externalities could include the presence of fines, restrictive contracts, subsidies, and taxes that the government could implement to re-balance the marginal private product with the marginal social product (Pigou, 1920). Knight (1924) criticized Pigou for does not agree with the concept of external economies and believed in market forces to appropriate the underlying economic effects of fluctuations in production costs. In this analytical framework, Knight believed in the free market to drive the economy to a higher level. No corrective actions to externalities were necessary.

For Daly and Farley (2010) Knight discussed in his article, "Risk, uncertainty and Profit" (1921), the case in which entrepreneurship supported the costs of failure and reaped the fruits of success, but in the case of exploitation of ecosystems often the entrepreneur appropriated the benefits and society bear the costs. A clear formation of negative externalities from the action of private agents oriented to profit. The interesting article by Ronald Coase, "The problem of social cost" (1960) recovered the discussion regarding externalities. For Coase, the problem of externalities should not have solely a marginal approach as Pigou pointed out, but also one regarding its global effects. Even finding a solution at the individual level to penalize agents for the negative effects generated, the correction could not produce the best results at the aggregate level: "The real problem to be solved is: A is allowed to damage B or B is right to harm A? The question is "how to avoid the most intense damage" (Coase, 1960).

Coase launches a new awareness solution, arguing that externalities are due to a lack of market and well-defined property rights. In his perception, the correct functioning of the market would ultimately lead to the final allocation of resources at the same level, even with externalities and regardless of whether the cause assumes or not its costs. For Coase, the Pigouvian solution of taxation, subsidies, or intervention does not necessarily raise the well-being of society. Coase further argues that one should seek the optimal solution because it believes that the benefits of removing the externality must be greater than the transaction costs to promote adjustment. That is to say that, not always the best solution is to tax the causer of the negative externality, the compensation for the loss, and disregard the market conditions in which are the involved agents. Direct negotiation or a clear definition of property rights could minimize transaction costs and increase the generated social benefit. On the other hand, when externalities achieve several agents that tend to infinity, the problem and the solution becomes complex.

Transaction costs can be prohibitive, interests and conflicts can be sharp, and agents cannot recognize the solution instances. When the costs tend to infinity and social benefits are more affordable than the cost of the transaction, the State becomes a key player in the resolution, following the Pigouvian line. Therefore, in more localized or well-established property rights disputes, the Coasean trading solution with low transaction costs can be an alternative to the internalization of externalities. In a larger arena, with systemic externalities, high transaction costs, and possible strategic actions, solely the performance of supra-individual representation like the State can reduce the externalities with an acceptable transaction cost.

### 3. Definitions of the microeconomics

Microeconomics textbooks reserve to externalities a section regarding "market failures". In said section, along with information asymmetries, public goods, incomplete markets, and market power, externalities are resorted to answering to the "deviations" of the neoclassical competitive market. Its definition is often both simplistic and contextualized to exemplify the effects of an external benefit or an external cost. Hunt (1980) E. K. Hunt, defines an externality, as follows:

> Whenever the usefulness for an individual is not a purely personal matter, individual, i.e., whenever the utility for a person is affected by the consumption of others (or by the production of firms), these interpersonal effects are referred to as "externalities".

Hunt's definition associates the rise of the debate regarding externalities with the main lesson of utilitarianism. The pursuit of self-interest ends up affecting other economic agents, even in an unnoticed way. The definition is generic and fits both the positive and negative externalities. Contemporary authors continue to reproduce externalities as a market failure in its two opposite sides:

> Externalities can arise between either producer, consumers, or even between consumers and producers. There are negative externalities - that occur when the action of one party imposes costs on another - and positive externalities - that arise when the action of one party benefits another. (…) An externality is an action by which a producer or a consumer affects other producers or consumers but does not suffer the consequences regarding the market price (Pindyck, 2002).

On the other hand, Mankiw (2009), also defines externality within the same reasoning:

> An externality arises when a person engages in an action that causes an impact on the well-being of a third party that does not participate in this action, without paying or receiving any compensation for said impact. If the impact on the third person is adverse, it is called a negative externality. If it is beneficial, it is called a positive externality.

Kreps (1990)Kreps strives to conceptualize externality in a wider way:

> the idea in a production externality is that a firm, by its choice of a production plan (or, for that matter, a consumer by her choice of consumption bundle) changes the feasible set of production plans for other firms.

Mas-Colell, Whinston, and Green (1990) define externalities, as follows:

> (…) we assumed that the preferences of a consumer were defined solely over the set of goods that she might herself decide to consume. Similarly, the production of a firm depended only on its own input choices. However, a consumer or firm may in some circumstances be directly affected by the actions of other agents in the economy; that is, there may be *external effects* from the activities of other consumers or firms.

Daly and Farley (2010), even within a heterodox structure, also define externalities:

> An externality occurs when an activity or transaction by some parts causes a loss or involuntary gain in welfare elsewhere and does not result in a compensation for the altering in welfare. If the externality results in a welfare loss, it is a negative externality, and if it results again, it is positive. The marginal external cost is the cost to society of the negative externality, that results from another "unit" of activity by the agent.

When externalities affect future generations, Daly and Farley (2010) warn of the possibility of infinite transaction costs between generations, and the lack of market self-correcting mechanisms without exogenous aid to the pricing system. To add to the former problem there is also imperfect information and uncertainty, making it difficult to equate marginal costs and marginal benefits regarding the likelihood of their behavior. The striking difference of Daly and Farley's approach is their long-term perception, being broader than the neoclassical framework that seeks the short-term balance. In microeconomics, there are also solutions to negative externalities. One of the principles usually adopted is the "polluter pays", where those who create the external cost take on the responsibility of negative effects to third parties. In this context, externalities can be mitigated without the presence of the State, if the transaction costs are not prohibitive, and the ones involved can reach an agreement. It is common to find the term "Coasean solution" (Mankiw, 2009).

Systemic externalities solutions typically involve the State, which includes linear taxation about the pollution level, taxing by pollution quotas, maximum emission quotas, tradable emission quotas, subsidies, fines, bans, and zoning. There is a mix of policies involving economic instruments and command and control instruments. In addition, the State can implement instruments of social communication, when it does not intend to tax or regulate too much a sector or a commodity good, but to highlight the consequences of consumption on the formation of negative externalities. It is common to find the term "Pigouvian solution" Mankiw, (2009). It is important to point out that, asymmetric information may cloak an externality, sending biased signals to the market, as it is the case of products that produce negative externalities during their respective production processes to which their consumers are unaware. In such a context, consumers make purchasing decisions that would not be taken if there was more complete information. Certification schemes are efficient options to signal to consumers that certain firms or production processes are internalizing externalities, fully or partially.

Another situation occurs when society does not in still altruistic and empathetic values. That is to say that, even with sufficient information on the externalities affecting society, consumer groups decide to maintain the same level of consumption, and in doing so, not affecting the market for that product generator of externalities. In general, this phenomenon occurs when the most severe effects of an externality occur in a distant location from the one in which the product is being consumed, and the benefits and costs are spatially separated (Alier and Jusmet, 2000). Interestingly in microeconomics is very used by authors, the expression "invisible hand" consecrated by Adam Smith. The autoregulation promoted by pricing and self-interest would be "a low-cost decentralized governance system." Daly and Farley (2010) criticize allocation in deregulated markets that generate results inefficient where "instead of individual self-interest by creating an invisible hand that maximizes social welfare, the marketing of goods, not the market creates an invisible foot that can kick the common good in the ass". Thus, in the economic textbooks there is a reference to the counterpart of the invisible hand, which draws attention as a process defined by Hunt (1980) as "invisible foot".

If the invisible hand organizes the social system, the invisible foot derails it. The two are antitheses of the same process, the search for personal satisfaction and the interlaced generation of benefits and costs. For Hunt, the treatment of externalities is the Achilles heel of the economy of well-being. Hunt assumes that an externality occurs when the value of a consumer's utility function is affected by the utility function of another consumer, or a firm's production function is affected by another firm's production function, or more importantly, that the individual function of an agent is affected by other agents not directly related. For conventional economics, except for an isolated externality, the Pareto optimality can be achieved. However, Hunt does not conceive the condition of subsidizing or taxing an isolated externality, because it would have a zero or neutral effect, as it could disregard millions of externalities that were occurring at the same time. The solution of an externality would not result in the solution of the Pareto optimal (Hunt, 1980). For Hunt, a large amount is not to say all the agent's decisions affect, to some extent, the pleasure or happiness of other agents. Only an individual extremist approach could accept an economic theory that considers the hypothesis of isolated externalities. Thus, the Pigouvian tax subsidy solution would be a fantasy, given the complexity of the presence of externalities. The solution would require millions of taxes and subsidies and the system would have endless rounds to equalize the internalization and would not take the system near the Pareto efficiency (Hunt, 1980).

Regardless, in Hunt's view, the most reactionary faction of orthodox economists occurred in the Coasean approach when it was no longer possible to ignore the environmental degradation in the United States at the end of the 1950s and the beginning of 1960. The policy of creating property rights to pollute the environment and then trade freely such rights was the tonic. The question was only a lack of property rights, and when this dispute was deemed resolved, the logic of maximizing benefits would lead the system to an efficient allocation of resources and the "optimal pollution". The Pareto optimality could be achieved, but with some level of environmental pollution. In this case, despite the economic equalization, nothing guarantees that the pollution levels are "optimal" for nature.

In his analysis, Hunt criticizes that such a solution would be able to resolve the issue of externalities. Hunt considers that in a model where the agent maximizes its benefits, the State establishes property rights, resulting always in the creation of new external diseconomies, each agent quickly realize that their decisions may impose external costs to third parties, so trading in the pollution market always leaves them in a better off situation. Being able to outsource portions of its costs to society, the agent does not restrain himself and creates the maximum social cost imposable to others. This would be Hunt's "invisible foot", which means that the within the system:

> The "invisible foot" ensures us that in a free-market, capitalist economy each person pursuing only his goodwill automatically, and most efficiently, does his part in maximizing the public misery.

Agents will choose systems capable of externalizing the maximum cost and guaranteeing net private benefits, kicking with their "invisible foot" as much of their responsibility as possible. The economy would be efficient in promoting misery (Hunt, 1980). Thus, the incentives to promote external economies would be weaker than the incentives to promote the lack of effectiveness in the use of productive means, causing an increase in the cost of production (that is external diseconomies). Neoclassical solutions of taxation, subsidies, and property rights would be inefficient. Thus as the orthodox approach is unable to analyse interdependent social forces, delimit property rights in the fields of physics and biology, and foster a rational taxation system that eliminates the ineffectiveness of the use of productive means (Hunt, 1980). However, Hunt's position is in line with that of Garret Hardin (1968), in the discussion of private benefit maximization and cost minimization, through the cost-sharing process. For Hardin, the "tragedy of the commons" is based on free access and unrestricted demand for a finite resource, which ends up structurally condemning the resource on account of its overexploitation. Each agent seeks to decide in their favours, but every decision involves a shared cost. If the agent adds a productive unit in his favour, but the cost is shared by the collectively, rationally there would be no impediment to such action. But, at the limit, the resources reach depletion and everyone ends up in a worse situation than the initial one. Hardin (1968) was criticized, as he was considered an apologist for the privatization of common resources and the suppression of collective ancestral lands, such as indigenous and marginalized populations in underdeveloped countries.

For Daly and Farley (2010), the market economy is a basic allocation institution, but society does not have institutions to limit the scale of the economy, especially beyond a point where the ineffectiveness of the use of productivity becomes the rule. Alier and Jusmet (2000) also discuss the problem of the generation of externalities, either in a strategically or randomly form with an analogy to the 'invisible elbow':

> When third parties not involved at all in an economic transaction, are affected by individual economic decisions, we could refer to the "invisible hand", to use the comparison of Jacobs in his book *The Green Economy*: when one moves to seek their interests, giving blows to others in a movement that sometimes he can be very aware of (as when a company is a clear focus of local pollution), but which often is unconscious (as when the decision of an individual contributes minimally, almost imperceptibly, to aggravate a global environmental problem).

The definition of "invisible elbow" is the action of pushing to third parties' part of their private costs, in an action that can be deliberate by the economic agent or be naive, in the sense that it is not realized its negative effects to third party agents. Whilst Elinor Ostrom revisited the tragedy of the commons from another perspective, positioning herself against the only regulation solution (Pigouvian solution) or the privatization of common resources (Coasean solution). Ostrom, (1990), appealed to the game theory and so, managed to understand the collective arrangements that create contracts of shared and responsible use of common resources. The self-interest of those involved in the use of a resource moves them to build contracts, often informal, to monitor and report violations of other agents and ensure contract compliance.

Indeed, the external costs approaches are controversial and there is no consensus regarding their solution to the systemic level. Conventional economic theory is conditioned on the definition that there are would-be market failures, and adjustments in specific sectors could lead to an efficient allocation. Hunt rejects this position and criticizes the structure of rational maximization of benefit, which would be most detrimental to the community.

Ostrom identifies solutions to common resources, but from collective actions and well-defined spaces. The debate continues and new analytical categories need to incorporate decentralized solutions at the local level but have systemic folding, low transaction cost, inclusive of the social point of view, reduced opportunistic strategic action, and that integrate economic concepts with the biophysical limits of ecosystems. Such integration could occur with the approach of economics with ecology, accepting mergers of sociology, anthropology, political science, geography, history, physics, and other sciences to form an interdisciplinary approach.

## 4. A new concept for the externality and the ecological cycle

In nature, there is a "functional division" of tasks. Solely 0.06% of the primary solar energy is converted by autotrophs into matter and chemical-biological energy, from its synthesis capability (Altvater, 1995). The base of the Earth's biomass depends on this conversion, including soil and climatic conditions adjusted to the limits of survival of plant species. Once solar energy is transformed into plant biomass, each growing cycle, the herbivores may prey in nature or receive power supplied by anthropogenic systems. Herbivores can be preyed upon or provided to carnivores following the food chain. Omnivores are at a level of consumption that includes feeding on autotrophs, herbivores, and any other types of biomasses. After the life cycle of autotrophs, herbivores, carnivores, and omnivores, a mass of energy and waste materials and structures that cannot be used directly by these mentioned species is formed. At this point, the "service" of decomposers, fungi, and bacteria enters. These microscopic species use dead matter as a substrate for their colonization and reproduction, being their food base. By using waste as raw material, they degrade macroparticles into smaller structures and release gases and water back into the environment. Its service is to recycle nutrients and make them available again for biogeochemical cycles. After transforming waste into micronutrients, gases in the atmosphere, and water in the various compartments, the autotrophs would be able to use such structures to restart the process of chemical transformation of solar energy into biomass. If autotrophs are the beginning of biomass formation, decomposers are the end of the cycle, but at the same time, it is the bacteria and fungi that ensure that there is no accumulation of waste and dead structures blocking the availability of nutrients for the life cycle to continue.

In this context, decomposers are reversing structures that, in principle, no longer have any utility. Animal/vegetable scraps and animal waste are, at the same time, the end and the beginning of the life cycle, being reversed from matter and disorganized energy (entropy) by the decomposers to become raw material again. The problem similarly is that this process is not clear in economic and human systems for sustainability to be a more concrete process (Daly 1991). In this case, there is a need to treat waste as if it were raw materials, or as if it were inserting a "decomposing function" in society. Namely, the "social autotrophic function" by Daly. This function has been studied since the classics, in the process of value creation. The other social "functions" are also clearer to be perceived. For example, the social functions that are perceived by the economics of the distribution of resources, or "who is the predator of whom", are objectively clear in economic theory.

However, the decomposing function performed by society as a service of dematerialization of the structures transformed by the economy into consumer goods is neither recognized nor valued as a service performed by social metabolism (Daly 1991). In this sense, the "decomposer function" is often not recognized or valued, for example, actions such as the service of decomposers in nature, as they are microscopic and difficult to visualize. Thus, it is believed that there are groups in society that also perform the decomposition service. They partially prevent the accumulation of scrap and waste on the surface of the economic, social, and environmental system. These groups of agents operate by reversing devalued or low-value macrostructures and making them available again to "autotrophic systems" that aggregate greater economic value. For example, companies and workers that add value to the garbage and/or waste by recovering them, thus avoiding a bigger problem with landfills. The initial reversibility refers to the action of identified diffuse economic agents that, operating in the market to pursue a monetary surplus in the self-interest of rationality transform the systemic entropy into the organized matter and energy again. The idea of reversal of entropy already has been discussed by (Boulding, 1966), (Georgescu-Roegen, 1971), (Daly, 1977), (Altvater, 1995) and many others, but what is added to this proposal is the reversal of negative externalities from the recycling of materials partially devalued from the material and energy point of view by agents other than those causing pollution.

In other words, firstly, entropy is reversed using appropriate technologies. If the system is operating above its support capacity, the action of the recycling agents becomes important to reverse this situation and bring the use of the resource back to the sustainable limit, even if the polluting agents and recyclers do not have perfect information about the sustainable limit of the system in operation. However, this reversal or throughput by Daly (1991), also includes the reduction of external costs not initially perceived by recycling agents, but intrinsic to diffuse agents. Since society partially or diffusely assumes these costs, the deliberate action of also diffuse agents can attenuate the formation of new social costs and their accumulation in the form of liabilities. At this point, the reversibility of the negative externality begins, here called a reverse externality. By advancing material recovery rates, it is possible to reach a point of stabilization of environmental liabilities, where recycling becomes circular production in a steady state, with the reversibility limit being an alternative solution for external costs. In an even more interesting position, the action of recyclers could still reduce accumulated liabilities tending to zero and could even create environmental assets that could generate diffuse marginal benefits to society.

The uniqueness of the current situation of the waste sector as an example is particularly interesting due to the destination of waste in economic production (re-entry of resources) and because part of the business community recently popularized the concept of "Circular Economy". We argue that it is correct to just call it circular production. Our caution regarding the term "Circular Economy" is notably due to the importance of academia, policymakers, and companies, which in this case we believe that the concept is still not very clear as to its definition, within the terms of economics. In this case, if there are advances in recycling rates, the objective of stabilizing environmental liabilities could be achieved, with the recovery of materials being a circular movement, as mentioned reversal limit would become a solution for external costs. In this way, the activities of scavengers and industrial firms are complementary, reducing environmental liabilities and generating external marginal benefits shared across society as a counterpoint to external social marginal costs, whose microeconomic impacts have an impact on the macroeconomic level of economic activity. But what is a "reverse externality" (hidden)? This new type of externality could potentially be occulting. For example, a green sector (the waste sector is unique) because can solve the negative externality in the short term, and therefore, it is more likely to identify a hidden externality.

*However, economists can argue that this sector by minimizing or mitigating a certain amount of negative externality in the green sector in question, is generating a positive externality. It is at this point that we consider the identification of "outliers". An "outlier" in this perspective presents a certain distance from the other points on the curve and the measurement of the distance, in this case, it has the property to identify anomalies "hidden externality" in question. We call this phenomenon an intrinsic point of border externality. Therefore, if the "outlier" is captured among the set of points under analysis, then we can understand that the unperceived (hidden) externality can have been identified. In this case, both short, medium, and long-term benefits are also identified and measured. In this perspective, theoretically, the hidden externality can serve to fill a part of the inefficiency gap of the market failure. (e.g., the reverse of the externalities).*

**Thus, the reverse externality can reveal itself as follows:**

a) Stationary: when the recovery rate of matter and energy equals the growth rate of demand for materials, keeping the extraction levels on the same level; this is the point where the stationary state of Daly is achieved (1991):

   a. Exhaustible: when the recovery of materials is the equivalent to the generation of waste, increasing the usage time of deposits in the long run, as it contains the short-term pollution and saves resources used in the extraction.

   b. Renewable: when recovery stabilizes the extraction of biogeochemical sources with renewal capability; it does not guarantee that the system is operating in the supported capacity but prevents the expansion of resources extraction.

b) Contractionary: when the recovery rate of matter and energy exceeds the demand growth rate, containing resources extraction levels:

   a. Exhaustible: when the recovery of scrap surpasses the generation of waste, increasing significantly the time of use of deposits in the long run, as it contains extraction.

   b. Renewable: when recovery stabilizes the extraction of biogeochemical sources with renewal capability, increasing the stocks of biological resources; it does not guarantee that the system is operating in the supported capacity, but it reduces the extraction of resources and promotes the formation of natural capital.

c) Mitigation: when the recovery rate of matter and energy is lower than the demand growth rate, not preventing the increase of extraction, but containing part of the potential resource extraction feed:

   a. Exhaustible: when the recovery of scrap is below the generation of waste, but it strongly prevents the time usage of deposits to be brought to the

> short-term, contains the short-term pollution, and saves resources used in extraction.

   b. Renewable: when the recovery does not prevent the increase in the extraction of biogeochemical sources with capacity for renewal but prevents its expansion more vigorously; it does not guarantee that the system is operating in the supported capacity, but prevents the expansion of resources extraction, containing at least partially the loss of natural capital.

It is important to point out that, the stationary state of Daly (1977) was criticized by Georgescu-Roegen (1971), since, even considering all materials being reprocessed to serve again as inputs to the economy, it is not possible to control the formation of entropy. It would be ecological salvation if such an entropic process could be blocked. In this way, what is being proposed as a reverse externality is more related to issues of economic order than material? The reversal of some or all the discarded material as waste would have an impact on reducing the flows of economic costs associated with depleted matter and energy, but it would be impossible to control all the entropy generated. Thus, following the Georgescu-Roegen logic, the proposal would not be to create the perpetual motorcycle, but to allow an expansion of the time of use of the materials and to avoid that the economic system reaches the limit of the optimal scale proposed by Daly and Farley (2010), where the marginal utility generated by the use of resources equated the marginal usability of these materials (the external cost materialized in the invisible foot or elbow).

The social decomposers would act freely in the market allocating their resources, seeking their self-interest, reducing the external costs even without strategic perception of this action. In this context, they could act on the "frontier" of the optimal scale of natural resource usage, reducing the speed of the economy to reach a certain scalar limit, breaking the system, or even pulling material savings back to a distant and secure point concerning equality between utility and marginal disutility of economic growth. This approach is a derivation of the classical solutions of externalities and maintains six perceptible differences. The first difference from the analytical models of Coase (1960) and (Pigou, 1920) refers to the substantial difference between the agents involved and the solution. Coase identifies polluting and impaired agents and constructs a solution of externality internalization by private property owners, lacking strategic action, infinitesimal transaction costs, and complete information. The Pigouvian model seeks to internalize externalities through the taxation of diffuse agents, known as polluters, accepting possible strategic action and considerable transaction costs. In our proposal for the reverse externality, we believe that diffuse private agents are operating in markets, competitive or oligopolistic, which end up reducing the external costs of other polluting agents, internalizing the operation of recycling and recovery systems.

The substantial difference is that in the reverse externality the internalization of cost does not occur by the polluting agent and is not derived from a polluter-impaired negotiation or even from a taxation levy. The action is free and is based on market forces, notably in signalling prices, in the business opportunity, or the possibility of social reproduction. By freely operating in the market, relying on waste and scrap from various production systems, now seen as throughput, private agents reduce the entropy and short- and long-term social costs that would be accumulating without the presence of these social recyclers.

Instead of accumulating entropy in pockets of waste, the operation of diffuse recycler systems would mitigate the logistic costs of collecting, the need for new spaces to land or dump waste, the control portion of the pollution caused, and reduce the need for new virgin raw materials. Thus, our understanding of reverse externality would be the process that derives from the action of private agents operating in markets for the recovery or recycling of materials and energy, seeking self-interest and based on the rationale of the opportunity cost of the capital or its social reproduction, which would be reducing costs and/or generating fuzzy benefits unobtainable in their gross revenue records or the cost structure of other fuzzy agents. In a more enlightening way, the benefits generated to other agents by the reduction of pollution or contamination would not be directly captured by the recycling agent, which would be a clear generation of external benefits, but not impediments to the maintenance of the recycling effort[1]. In addition, the reduction of the external cost of operation of other agents would not be directly perceived by the recycling agent, but its action directly contributes to contain the social cost of pollution, contamination, and scarcity of virgin raw material.

The second difference in approach is that the reverse externality differs from the negative externality because, in the first, agents recognize entropy as a business opportunity, while in the second approach agents recognize entropy as a cost to be assumed both in their treatment and in their compensation to impaired agents. In the negative externality, the process is an end of the pipe, an "invisible foot/elbow", where agents seek to get rid of the problem of pollution by trying to outsource to society, raising the social cost. In the reverse externality, agents reduce the social cost by using waste/scrap as a source of raw material for their enterprises, controlling the formation of entropy with clear refutations in other productive systems and social life. In the presence of pollutants, operating freely in the market, and considering the waste of other enterprises as a throughput, one can visualize the possibility of shifting the marginal cost curve of pollution to the left, reducing the optimal pollution, but maintaining the level of social production of goods and services, without loss of well-being. In addition to reducing the entropy in the system, the recyclers promote the saving of materials not yet exploited, extending the time of use of the deposits or renewable resources. It may not represent a solution to the generation of entropy, but the presence of these agents leads to time gains until another socio-environmental institutional arrangement is established as a new paradigm.

Therefore, just as microscopic as it is, the ecosystem service of decomposers in nature, it is also difficult to visualize this same type of service in the economic system. However, the initiatives of collector workers (diffuse) and recycled industries (identified) are paradoxical for the perception of value and usefulness of individuals and societies. Both economic agents assume the external costs of those causing the negative externality. The first initiative is intrinsic to informal markets in developing countries, while the second initiative takes place in the formal market in developed countries. In this case, the so-called "scavengers" and "decomposers" perform the economic ecosystem service of decomposing the material structures of the economy, partially avoiding the accumulation of leftovers and residues on the surface of the economy, social and environmental system (Ghisellini et al., 2016). These individuals and firms have, intrinsic to their operations of activities, impacting effects that influence and modify the devalued macrostructures, making them again available to be used by aggregators of the greater market and economic power (Ghisellini et al., 2016; Geng & Côté, 2002). Figure 1. Show Biophysical thresholds of expansion of the economic system, the utility- and disutility marginal relationship.

---

[1] In the traditional approach, the generation of external benefits creates a disincentive to economic agents for not being able to protect all marginal benefits in their utility function, with clear consequences of underinvestment, under-allocation of resources and loss of economic growth potential.

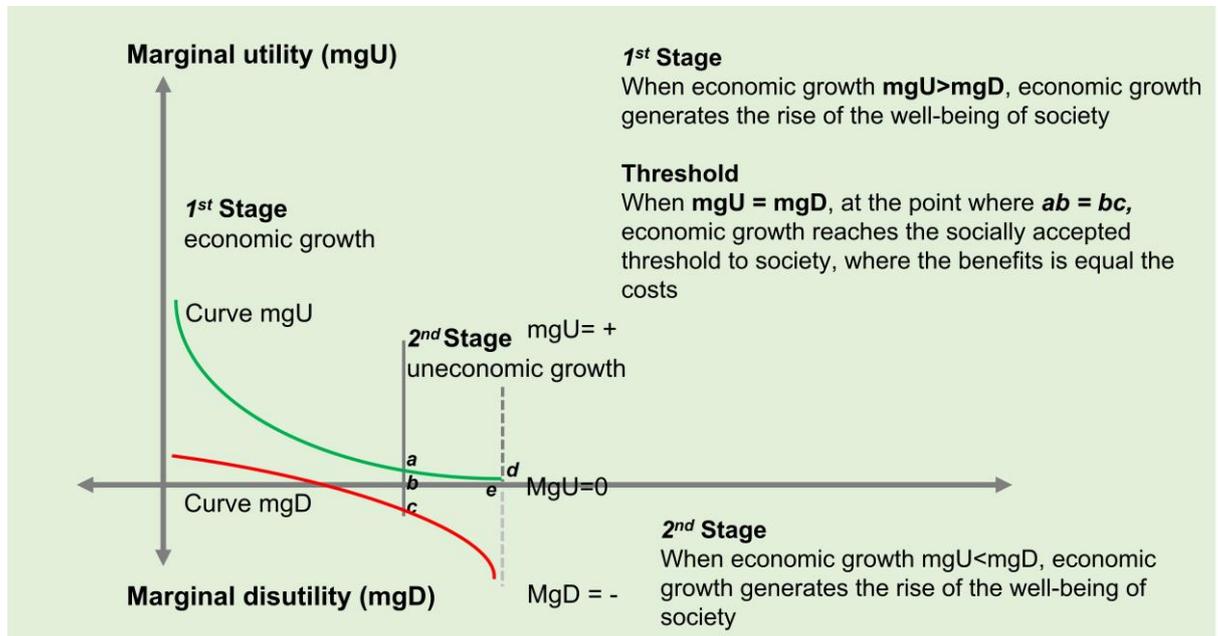

Figure 1. Biophysical thresholds of expansion of the economic system, the utility- and disutility marginal relationship (adapted of Daly and Farley, 2010).

Figure 1 explains how natural resources extracted from nature have marginal utility (we will discuss how this occurs in another article) and at the same time produce a negative externality (entropy), as they deplete them, which, as a result, produces disutility marginal part of natural resources. On the other hand, the transformation of waste into (enthalpy) productive resources has marginal utility. This result is due to the reversal of the disutility of natural resources. In other words, the utility of the positive externality for a given volume of energy extracted from waste is equal to the disutility of that same volume of waste generated.

This would be the third difference because, in the limit, the internalization of externality by the reverse action does not imply supply beyond the point of economic efficiency, as the external costs that were being imposed on society are reversed by private agents. This is an important improvement over the classic internalization solution, which implies reduced production and the adjustment of the production excess. As Daly and Farley (2010), the presence of negative externality producer agents lead to activity too far from where the point of Pareto would be efficient (Phase II in Figure 1), but with the presence of reverse externality, external costs are partially or fully absorbed by the productive system, preventing the solution to be cutting production, generating unemployment, and tax payment reduction (Phase I in Figure 1). Only if the production is above the ecological carrying capacity, it should be sought to reduce the scale of production and bring the use of energy-material to the sustainable level. Otherwise, if the product is still within the maximum use of resources, it would be necessary to reduce and loss of material production, but to ensure the presence of the reverse phenomenon externality. As observed in Figure 1, reverse externality could slow the economic system advance to phase II, decelerating or braking input at the phase of marginal disutility. In this case, the creation of taxation or negotiations striving that the polluter internalizes his externality is not enough to generate a reduction in production q' to q'. This is because part of the external cost would be reversed by recycling agents and the polluter would incur lower taxation or negotiation costs. Portions or all the marginal costs of the polluter pollution would be absorbed by recycling agents that recognize pollution as throughput and not as a cost.

The fourth difference in the externality approach is that in our proposal, reversibility starts as an ecological-economic approach because the first variable to be identified and reversed is entropy. Once perceived the reversal of the generated entropy, one proceeds to the integration with the economic variables that signal agents to market opportunity. In the Coasean-Pigouvian approaches, there is the concern of internalizing the external costs, trying to find the point of "optimal pollution", but it is not addressed the issue of entropy and the point limit of biophysical carrying capacity. Here it is clear that the most important action is to keep the energy-material system within operating safety limits and that the market can assist in balancing solutions between supply, syntropy, and entropy generation. Specifically, this theoretical model seeks to link the ecological-energy approaches to economic, to find balance in the use of natural resources and the efficient allocation of economic resources, while the traditional approach equalizes pollution only from the economic point of view.

The fifth difference refers to the importance of economic agents involved. In the presence of reverse externality, it becomes clear that recyclers are usually devalued social actors, on the margins of the economy, that do not have market power and are often informal. They operate in the base system, collecting scraps and discarded waste. A reverse externality can be associated with a vision of social basis, raising the importance of the small "excluded" ones, who work as social decomposers, sweeping the garbage, and recycling surface entropy of the economic and social metabolism, without the recognition of society. The action of recycling social groups is the equivalent to capital as a stabilizing system action, such as, raising the constant technological capital in response to the increasing entropy, which in this case has a similar function to recyclers. That is to say that one can equate the mitigating action of recyclers, even informally, at the same level of the social importance of productive capital. This debate is not included in the Coasean and Pigouvian solutions.

In addition, the sixth difference in the solution approaches would be about the collective arrangements of Ostrom. Here the solution is not to delimit the common *ex-ante* and formulate contracts, even if informal ones. The decentralized and individualized solution of economic agents does not imply a contractual arrangement. Social actors can operate alone and observing their self-interest, but when they position themselves in the "social-metabolic decomposing function", they end up generating positive effects on the social fabric and its surrounding ecosystems.

## 5. Between the invisible hand, elbow, and foot: using the rational mind

In the current structure, the work of recyclers would not have a "social value" equivalent to the productive and financial capital, because the natural capital it recovers (the natural capital scrap) is not measured correctly and so the work of recyclers is not recognized, often seen as "unnecessary", "undervalued", or "poor" (as street collectors). However, their actions recover portions of the surplus that was being deposited as waste in landfills and scrap sinks. The debate has not yet evolved enough to characterize how this kind of surplus is recovered. What would be this surplus value? The capitalist system appropriates the surplus labour, but after the production and circulation, the value recovered by recyclers of natural capital scrap still represents a surplus of work or a kind of residual natural capital. Can the recycler retain a portion of that surplus? Is the amount you receive part of the surplus recovered or only the value of the allocated work? Here arise fundamental issues that can subsidize assurance policies for the value of surplus recovered from scrap and waste, to reduce exploitation in value recovered by the action of recyclers. A working proposal would be the creation of a minimum package of incentives to strengthen the process and ensure value for recyclable goods as a similar system to guarantee agricultural commodity prices and minimum wage for producers and workers. It would be a socio-economic and environmental proposal as it would recover natural capital scrap, avoids the expansion of the exploitation of natural intact capital, generates surplus value, and distributes income to the most vulnerable social groups.

The expansion of the phenomenon of reverse externalities and their presence in a systemic way in a society can generate significant effects on an increasing scale. Individual agent's aggregating actions may rebut to the macroeconomic level. For example, the phenomenon of reverse externality generated by diffuse economic agents can reduce the operating cost of the collecting, treatment, and storage systems of waste which is taken care of by the community. The resources saved by the State can be allocated to other generating systems of a positive externality, such as education, science and technology, and other actions. In expanded form, reverse externality can reduce part of the pollution and the absolute and relative scarcity of natural resources and raw materials, reducing the gap between the marginal utility and the marginal uselessness of economic growth, considering the maximum range of use of natural resources. In this case, the reverse externality would occur to the point where marginal costs were reversed by recycling agents, bringing the system to the limit of sustainable capacity. From this point on, initiates the generation of marginal benefits.

Without the presence of social recyclers, the operating range of the system would be shortened, raising social action as the condition to maintain the current productive structure. Reverse externality could be stimulated in markets where it is possible to transform entropy in throughput, creating job opportunities, surplus generation, and reverberation in the control of entropy and in the expansion of the time of usage of natural capital. Otherwise, reverse externality could generate some extra time to find other institutional, social, and productive arrangements more sustainable and build more favourable scenarios to a widespread recycling system.

More than believing in the invisible hand with the creation of pollution markets or simply accept the invisible foot/elbow and the total failure of the system, we bet in the middle position. There are thousands of economic agents, social actors, or simply workers who seek their everyday solution for survival in the recoverable value of scrap, waste, and carcasses of the social metabolism. Their action is not altruistic, they do not operate with contractual arrangements, no strategic action, and they do not have property rights over the scraps. They collect what the invisible foot/elbow threw out the system, but rather than noticing the leftovers/waste as an externality, they perceive them as the source of a resource to achieve an economic surplus. The greater the strength of the "foot" and "elbow", the greater the value recovery possibilities. The orthodox solution and Ostrom do not include the reversibility of entropy through the work and efforts of millions of diffuse social recyclers. We can think of rejecting the invisible hand, elbow, and foot as established positions and propose the use of the visible head, and rational mind, something far more rational and ecological than solely believing in far more obscure processes with great difficulty of implementation.

6. Conclusion and proposal limitations

A weakness of the proposed technique is that at a social level there could be the formation of a lock-in process in the employed technology because the life expectancy of prime matter could be partially recovered and/or extended. In a scenario of absence or a lower rate of recycling agents, sources of natural resources could be quickly exhausted, creating pressures for technological innovations to overcome established ones. Also, the lack of raw materials could motivate the use of background technologies with a higher introduction of operative costs. That is to say that, in the context of multiple dynamic interactions, the presence of reverse externality can support the mitigation of short-term external costs, also they can block or delay the implementation of a new technological paradigm. Planners should be aware of possible technologies that may overcome a structure with large external costs generation requiring recycling agents to ensure that such ventures keep running.